\documentclass[preprint,12pt,review]{elsarticle}

\usepackage{amssymb}
\usepackage{amsmath}

\journal{arXiv.org}

\begin{document}

\begin{frontmatter}



\title{Production of heavy vector particles in interactions of leptons with the massless gauge bosons}


\author{I. Alikhanov\corref{cor1}}

%%Email address, Telephone and Fax numbers
\cortext[cor1]{{\it Email address: {\tt ialspbu@gmail.com}}}
%%%%%%%%%%%%%%%%%%%%%%%%%%%%%%%%%%%%%

%%%%%%%%%%%%%%%%%%%
\address{Institute for Nuclear Research of the Russian Academy of Sciences,
60-th October Anniversary pr. 7a, Moscow 117312, Russia}
%%%%%%%%%%%%%%%%%

\begin{abstract}
The cross section for vector leptoquark production in electron--gluon collisions is calculated analytically using the Lagrangian with the minimal couplings between the leptoquarks and the gauge fields of the standard model. It is found that the cross section significantly exceeds the corresponding quantity previously presented in the literature.  The cross section of exclusive $W$~boson production in neutrino--photon scattering emerges as a by-product of this letter. The obtained results can be used for studies at $ep$ colliders.  
\end{abstract}


\end{frontmatter}
%\maketitle %

%%%%%%%%%%%%%%%%%
\section{Introduction}
\label{intro}
The cross section being the most important quantity for description and interpretation of physics underlying interactions of particles connects theory and experiment. Accurate knowledge on it is therefore very important for deeper understanding already well established theories like the standard model as well as for search of new physics.

The effective Lagrangian of the leptoquark model proposed by Buchm\"uller, R\"uckl and Wyler in 1986~\cite{leptoquark} obeys the symmetries of the standard model gauge groups $SU(3)_C\times SU(2)_L\times U(1)_Y$. Therefore one may expect that  dynamics of some processes in these models can formally coincide so that the calculated cross sections turn out also to be the same up to constant factors associated with different couplings. 
This letter shows that exclusive production of $W$ bosons in neutrino--photon scattering and production of vector leptoquarks in electron--gluon collisions represent such a situation in the leading order of perturbation theory provided the leptoquarks minimally couple to the gauge fields of the standard model. The cross sections of both reactions are calculated analytically. A significant difference between the result of this letter and previous analysis of the leptoquark production is found.

%%%%%%%%%%%%%%%%%%%%%%%%%%%%%%%%%%%%%%%
\section{Comparison of the models}
\label{problem}
%%%%%%%%%%%%%%%%%%%%%%%%%%%%%%%%%%%%%%

Following~\cite{montalvo} let us also postulate the Lagrangian with minimal couplings between vector leptoquarks and the neutral gauge bosons of the standard model:

\begin{eqnarray}
\mathcal{L}^{\gamma,Z,\text{g}}=\sum_{V}\left[-\frac{1}{2}G^{\dagger}_{\mu\nu}G^{\mu\nu}+M^2_{V}V^{\mu\dagger}V_{\mu}\right.\\\nonumber\left.-i\sum_{j=\gamma,Z,\text{g}}g_jV^{\dagger}_{\mu}V_{\nu}F^{\mu\nu}_j\right],\label{lagrange}
\end{eqnarray}

where $G_{\mu\nu}=D_{\mu}V_{\nu}-D_{\nu}V_{\mu}$ is the field strength tensor of vector leptoquarks, $F_{\mu\nu j}$ are the field strength tensors of the neutral gauge fields of the standard model, $g_j$ are the coupling constants for the interactions between these gauge fields and the leptoquarks.
The covariant derivative $D_{\mu}$ is defined by

\begin{equation}
D_{\mu}=\partial_{\mu}-ieQA_{\mu}-ieQ^ZZ_{\mu}-ig_s\frac{\lambda^a}{2}A^a_{\mu}.\label{derivative}
\end{equation}

Here $A_{\mu}$, $Z_{\mu}$ and $A^a_{\mu}$ are the photon, $Z$ boson and gluon fields, respectively; $e$ is the elementary electric charge, $Q$ denotes the electric charge of a leptoquark, $g_s$ is the strong coupling constant, $Q^Z=(T_3-Q\sin^2\theta_W)/\sin\theta_W\cos\theta_W$ ($T_3$ is the third component of the weak isospin, $\theta_W$ is the Weinberg angle), $\lambda^a$ are the Gell-Mann matrices.

\begin{figure}
\centering
\resizebox{1.0\textwidth}{!}{%
\includegraphics{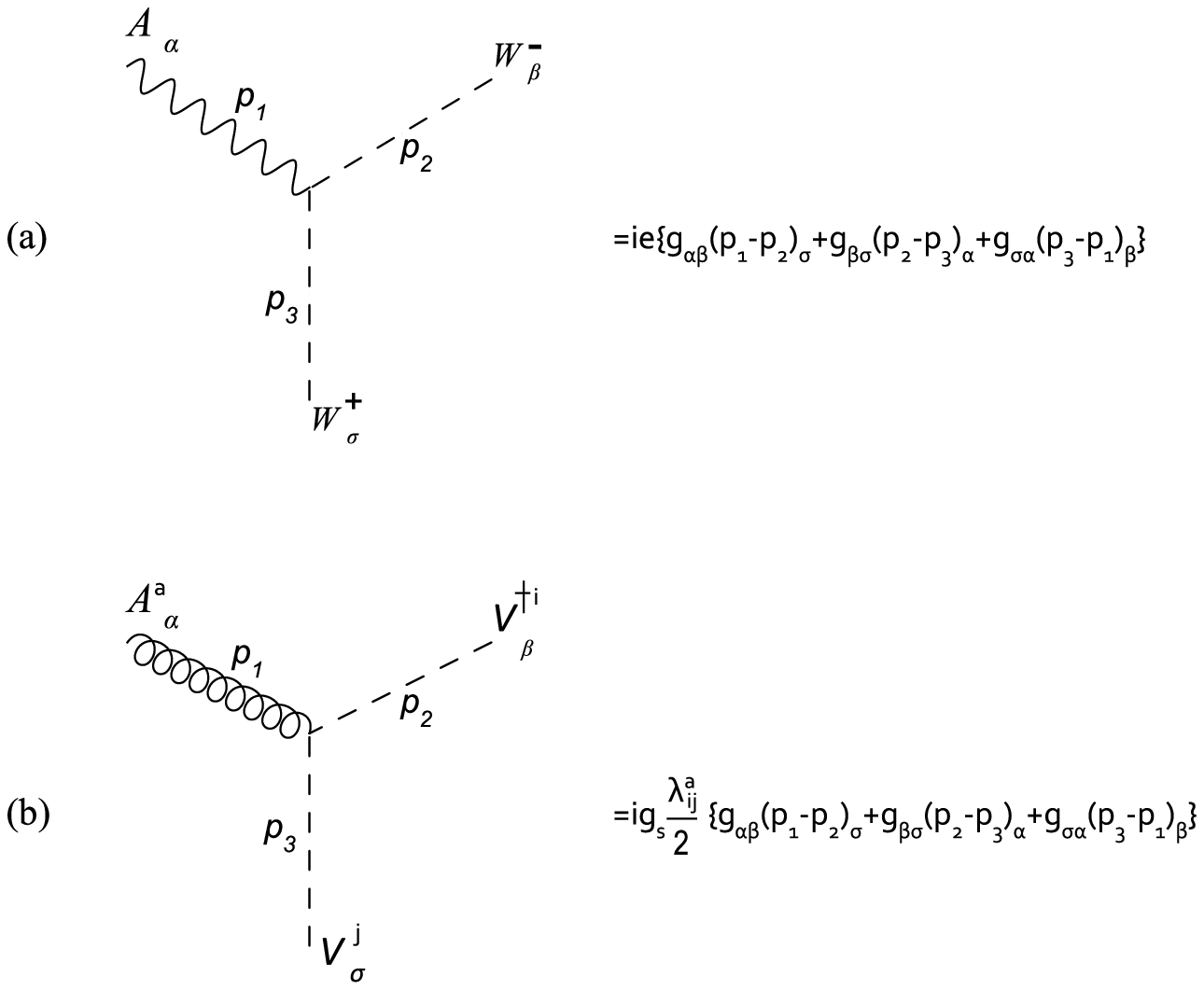}
}\caption{}
\label{fig1}
\end{figure}

From (\ref{lagrange}) it follows that the Feynman rules for the $ZVV$ and $\gamma VV$ vertices are similar to the $ZWW$ $\gamma WW$ ones~\cite{montalvo}. The couplings of vector leptoquarks to the gluon fields  have also analogy with the self-interaction of the gluons. For example, in Fig. \ref{fig1}  the explicit form of the Feynman rules for the $\gamma WW$ and $\text{g}VV$ vertices are given. One can see that they differ only by constant factors.  

Let us consider two exclusive reactions. The first one is the $W$~production in neutrino--photon scattering allowed by the standard model \cite{seckel}:

\begin{equation}
\nu_l+\gamma\rightarrow l+W,\label{reac1}
\end{equation}

where $l=$e$,\mu,\tau$.

The second one is production of vector leptoquarks ($V$) in interactions of left/right polarized electrons with gluons appearing in the Buchm\"uller--R\"uckl--Wyler leptoquark model~\cite{spira}:

\begin{equation}
e_{L/R}+\text{g}\rightarrow q+V.\label{reac2}
\end{equation}

\begin{figure}
\centering
\resizebox{1.0\textwidth}{!}{%
\includegraphics{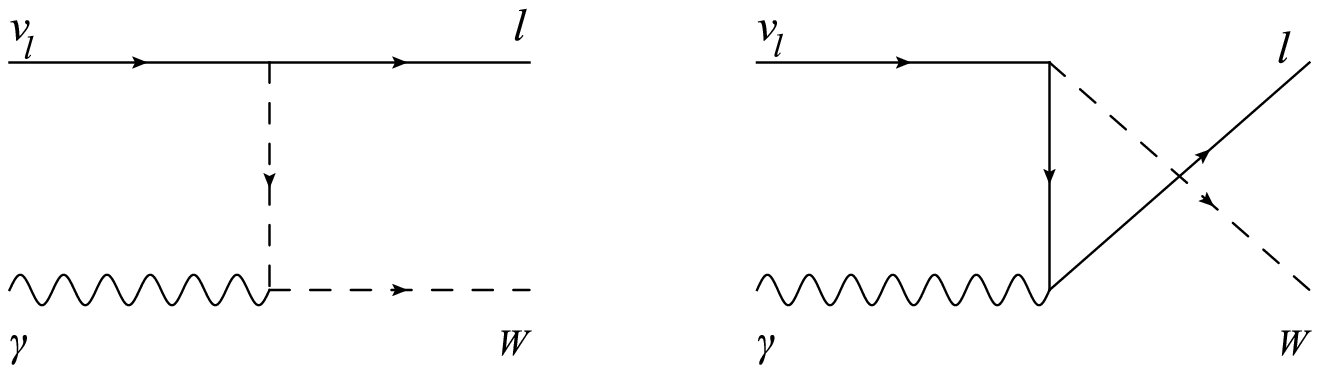}
}\caption{}
\label{fig2}
\end{figure}

\begin{figure}
\centering
\resizebox{1.0\textwidth}{!}{%
\includegraphics{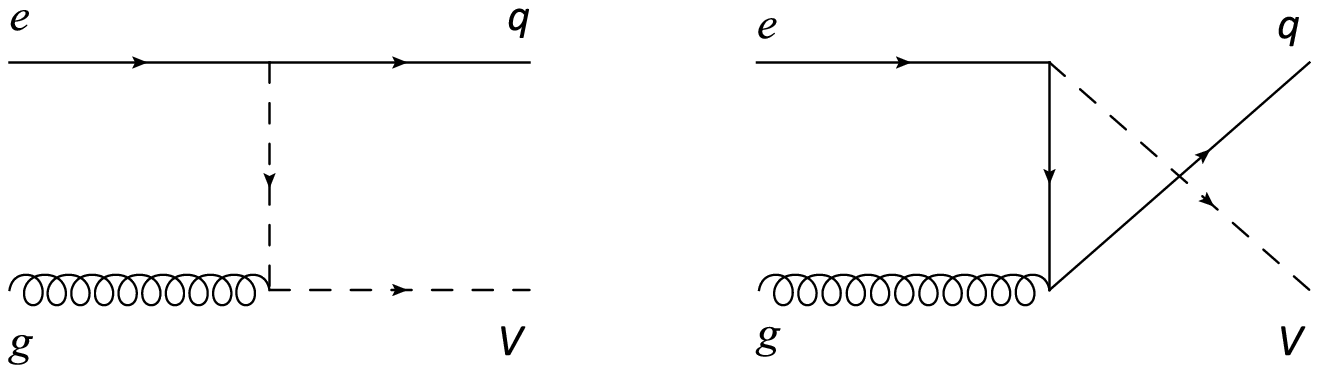}
}\caption{}
\label{fig3}
\end{figure}

The reactions (\ref{reac1}) and (\ref{reac2}) are closely related to each other from a formal point of view. To illustrate this, it is convenient to represent the considered interactions in the form of the Feynman diagrams so that the amplitudes contributing to (\ref{reac1}) and (\ref{reac2}) will look as shown in Fig. \ref{fig2}  and Fig. \ref{fig3}, respectively \cite{seckel,spira}.

\begin{figure}
\centering
\resizebox{0.9\textwidth}{!}{%
\includegraphics{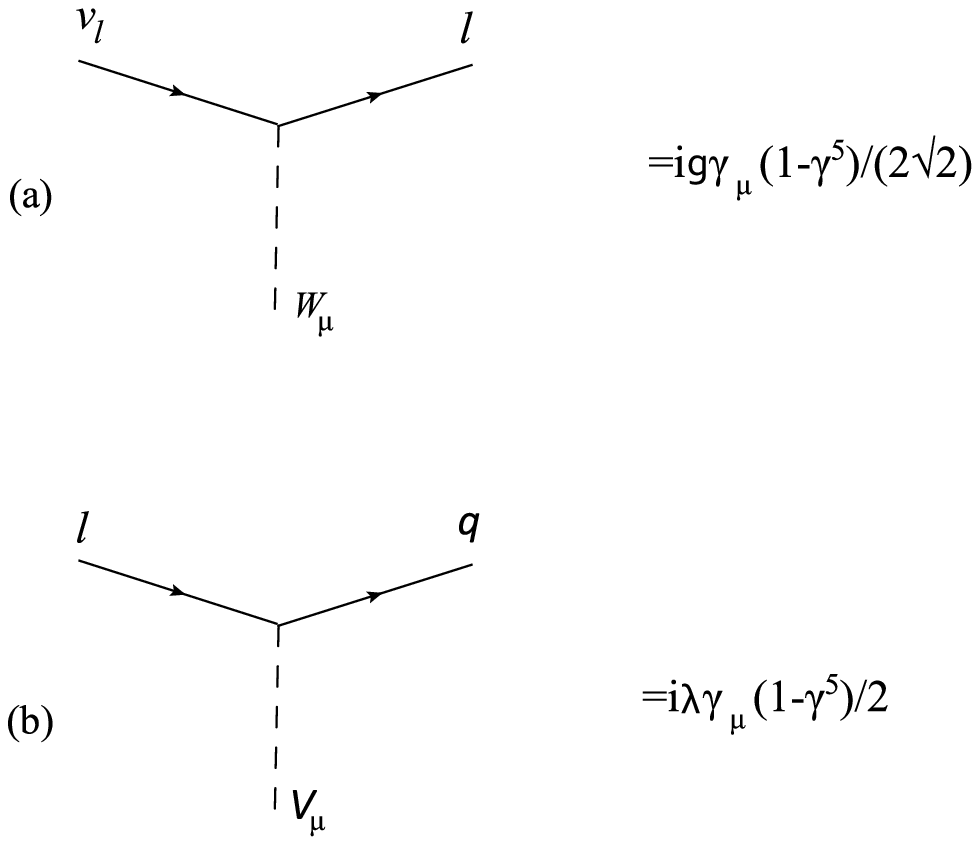}
}\caption{}
\label{fig44}
\end{figure}

The leptoquark Lagrangian has such symmetry properties that the Feynman rules for the vertices $Veq$ respectively coincide in their structure with those of the standard model for the vertices $W\nu l$ up to constant factors (see Fig. \ref{fig44}). As to the familiar rules corresponding to the $\gamma ll$ and $\text{g}qq$ vertices, they also differ from each other only by the coupling constants and the color coefficients. This means that in the limit of massless initial state leptons the leading order cross sections of the reactions (\ref{reac1}) and (\ref{reac2}) will have exactly the same dependence on the Mandelstam variables and on the masses of the final state particles differing from each other only by a constant as well. Such relations between processes are well known in quantum field theory. Most simple QCD diagrams are exactly analogous to QED diagrams, and the QCD cross section is obtainable by appropriate replacements of the coupling constants in the QED one corresponding to the replacement of the photon by a gluon~\cite{helzen}.

Suppose that $\sigma_1(s,M_W,m_l)$ and $\sigma_2(s,M_V,m_q)$ are the leading order cross sections of the reactions (\ref{reac1}) and (\ref{reac2}), where $s$ is the center-of-mass (cm) energy squared, $M_i$ and $m_j$ are the masses of the final state particles. Then, in accordance with the above discussion, they must satisfy the following condition:

\begin{equation}
\frac{\sigma_1(s,M,m)}{\sigma_2(s,M,m)}=\text{const.}.\label{ratio1}
\end{equation}

Note that both cross sections in (\ref{ratio1}) are taken with the same values of the masses.
Therefore, it is enough to know one of the cross sections to find the other.

%%%%%%%%%%%%%%%%%%%%%%%%%%%%%%%%%%%%%%%%%%%%%%%%%%%%%%%%%%%
\section{Calculation of the cross section\label{calcs}}
%%%%%%%%%%%%%%%%%%%%%%%%%%%%%%%%%%%%%%%%%%%%%%%%%%%%%%%%%%

The cross section of the reaction (\ref{reac2}) is calculated using the diagrams from Fig. \ref{fig3} with the corresponding Feynman rules given in Figs. \ref{fig1}(b) and \ref{fig44}(b). The result reads

\begin{eqnarray}
\sigma_{2}(s,M_V,m_q)=\lambda^2\alpha_s F(s,M_V,m_q),\label{result}
\end{eqnarray} 

where $\lambda$ is the coupling constant corresponding to the $Veq$ vertex, $\alpha_s=g^2_s/4\pi$,

\begin{eqnarray}
F(s,m_1,m_2)=\frac{1}{8m_1^2}\left\{\frac{p}{s^{5/2}} \left[8s^2+16m_1^4-11m_2^2\left(s+m_1^2\right)-5m_2^4\right]\right.\nonumber\\\left.+2\frac{\,2 m_1^2 \left(s^2-2m_1^2s+2m_1^4\right)+m_2^2\left(s^2-\left(s-3m_1^2\right)\left(m_1^2-2m_2^2\right)\right)-m_2^6}{s^3} \right.\nonumber\\
\left.\times\log
   \left[\frac{\left(p^2+m_2^2\right)^{1/2}+p}{m_2}\right]\right.\nonumber\\\left.-\frac{8 m_1^4\left(s-m_1^2\right)-m_2^2 \left(s^2+4m_1^2s+6m_1^4\right)-m_2^4 \left(s-12m_1^2\right)+2 m_2^6}{s^3}\right.\nonumber\\
\left.\times 
   \log   \left[\frac{\left(p^2+m_1^2\right)^{1/2}+p}{m_1}\right]\right\}.\nonumber\\\label{function}
\end{eqnarray}

Here $p=\sqrt{\left(s-(m_1+m_2)^2\right)\left(s-(m_1-m_2)^2\right)}/2\sqrt{s}$ is the cm momentum of any of the final state particles.

%%%%%%%%%%%%%%%%%%%%%%%%%%%%%%%%%%%%%%%%%%%%%%%%%%%%%%%%%%%
\section{Verification of the validity of the cross section}
%%%%%%%%%%%%%%%%%%%%%%%%%%%%%%%%%%%%%%%%%%%%%%%%%%%%%%%%%%

The similarity between the cross sections of the reactions (\ref{reac1}) and (\ref{reac2}) discussed in Section~\ref{problem} allows to verify the validity of the cross section~(\ref{result}). 
Actually, the problem of calculation of $\sigma_1(s,M_W,m_l)$ is now reduced to just performing the following obvious replacements of the coupling constants and the masses of the final state particles in (\ref{result}):

\begin{equation}
\lambda\rightarrow\frac{g}{\sqrt{2}},\: \frac{\alpha_s}{2}\rightarrow\alpha,\:M_V\rightarrow M_W,\:m_q\rightarrow m_l, \label{replace1}
\end{equation}

where $g$ is the coupling of the weak charged current (related to the Fermi coupling constant $G_F$ by $G_F=\sqrt{2}g^2/8M_W^2$, $\alpha$ is the fine structure constant. Note that the coefficient of $\alpha_s$ is the color factor equal to $1/2$  for the process~(\ref{reac2}).

So that one finds

\begin{eqnarray}
\sigma_{1}(s,M_W,m_l)=g^2\alpha F(s,M_W,m_l),\label{result2}
\end{eqnarray}

The cross section for the reaction (\ref{reac1}) in the leading order has also been independently calculated in \cite{seckel}, however in such a way that the masses of the final state leptons were neglected (let us denote this cross section by $\sigma^c_{1}(s,M_W,m_l)$). This means that (\ref{result2}) obtained in this letter and the result of~\cite{seckel} must asymptotically coincide satisfying the following condition: 

\begin{equation}
\frac{\sigma_1^{c}(s,M_W,m_l)}{\sigma_1(s,M_W,m_l)}\Biggl\rvert_{\sqrt{s}\gg M_W+m_l}\hskip -1.7cm=1.\label{ratio2}
\end{equation}

It should be emphasized that (\ref{ratio2}) is a criterion to verify the validity of the cross sections obtained in the present letter.

\begin{figure}
\centering
\resizebox{0.9\textwidth}{!}{%
\includegraphics{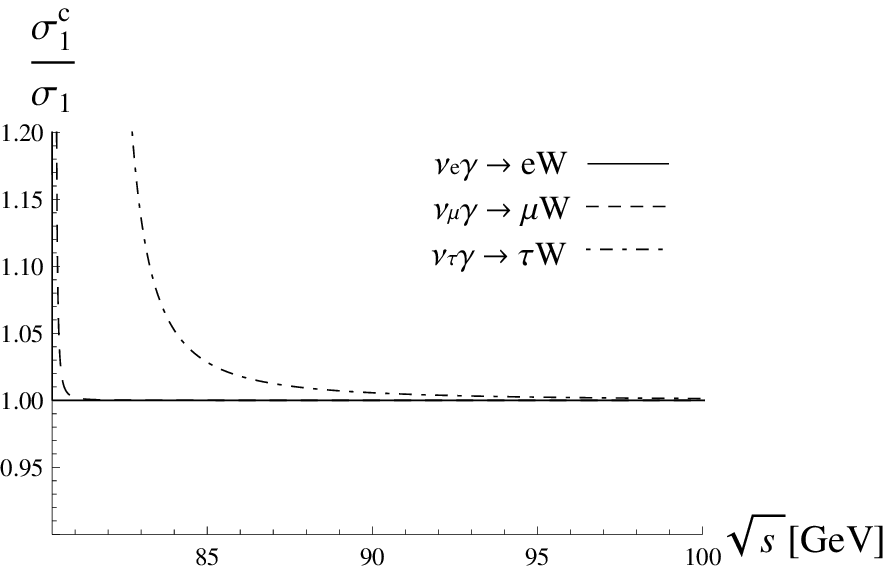}
}\caption{}
\label{fig55}
\end{figure}

Figure \ref{fig55} shows that (\ref{ratio2}) is fulfilled as it must be in the case of correct calculations. Moreover, there is a perfect coincidence between  (\ref{result2}) and $\sigma^c_{1}(s,M_W,m_l)$  over the entire range of the energy for the case of the electron in the final state owing to the vanishing electron mass in comparison with~$M_W$.

%%%%%%%%%%%%%%%%%%%%%%%%%%%%%%%%%%%%%%%%%%%%%%%%%%%%%%%%%%%
\section{Comparison with previous calculations}
%%%%%%%%%%%%%%%%%%%%%%%%%%%%%%%%%%%%%%%%%%%%%%%%%%%%%%%%%%

\begin{figure}
\centering
\resizebox{1.2\textwidth}{!}{%
\includegraphics{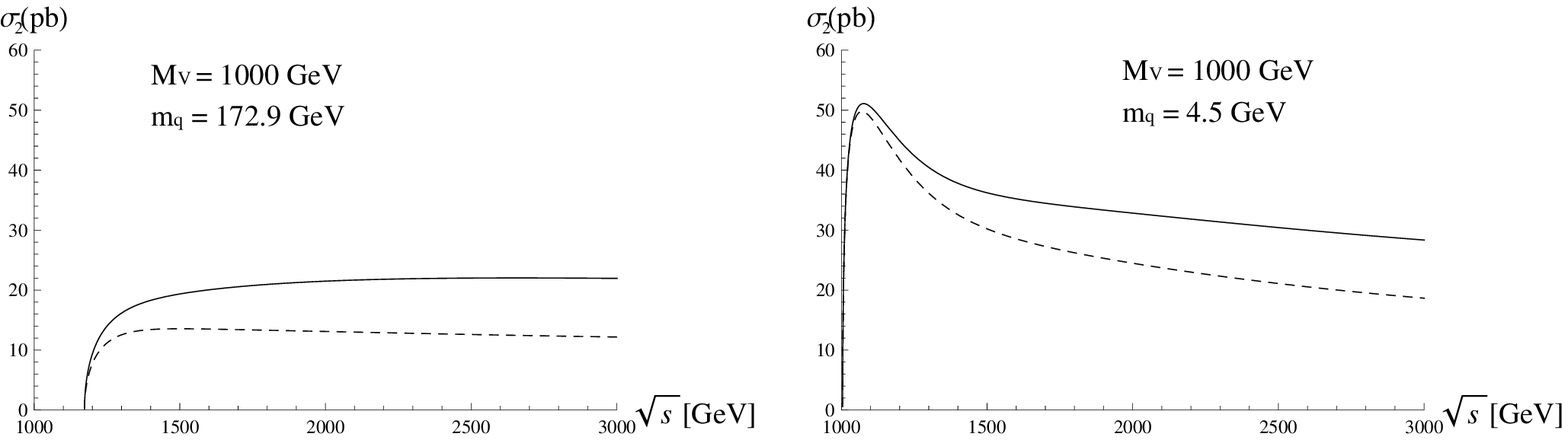}
}\caption{}
\label{fig5}
\end{figure}

\begin{figure}
\centering
\resizebox{0.9\textwidth}{!}{%
\includegraphics{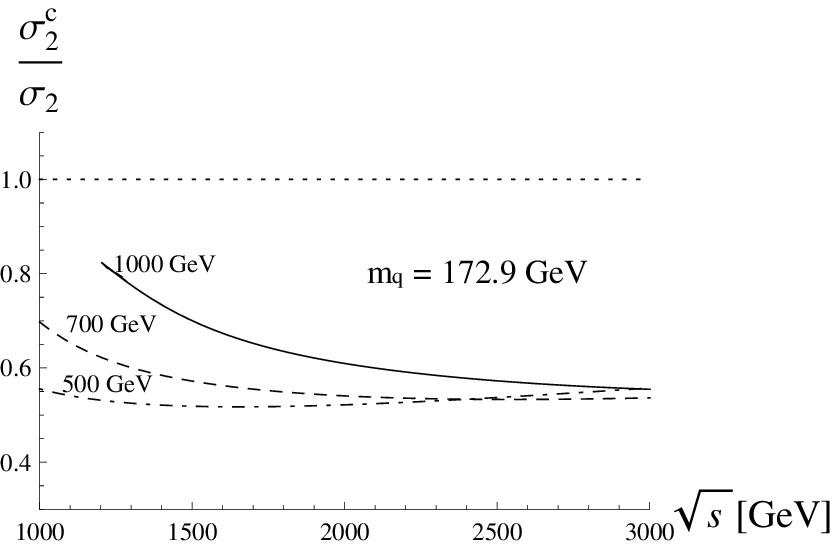}
}\caption{}
\label{fig6}
\end{figure}

The cross section for the reaction (\ref{reac2}) in the leading order has also been calculated in \cite{spira} whose result (let us denote it by $\sigma^c_2(s,M_V,m_q)$)  significantly differs from (\ref{result}). This fact is illustrated in Fig.~\ref{fig5}, where dependences of both cross sections on the cm energy are shown for production of the $(eb)$ and $(et)$ type vector leptoquarks of mass 1000 GeV at $\lambda=1$ and $\alpha_s=0.118$. The difference is more brightly reflected by the ratio of $\sigma^c_2(s,M_V,m_q)$ to $\sigma_2(s,M_V,m_q)$ shown in Fig.~\ref{fig6}. One can see that there may be cases in which the cross section (\ref{result}) exceeds the corresponding quantity from \cite{spira} by about a factor of two. This is because different choices of the Lagrangian responsible for interaction of the leptoquarks with the gluon fields which lead to different Feynman rules for the $\text{g}VV$ vertex \cite{private}. In the present letter the minimal couplings of the leptoquarks to the gauge fields of the standard model are assumed while in~\cite{spira} the Lagrangian with anomalous interaction terms is used~\cite{vertex, vertex2}. 

%%%%%%%%%%%%%%%%%%%%%%%%%%%%%%%%%%%%%%%%%%%%%%%%%%%%%%%%%%%%%%%%%%%%%%%%%%%%
\section{Production of vector leptoquarks in electron--nucleon collisions}
%%%%%%%%%%%%%%%%%%%%%%%%%%%%%%%%%%%%%%%%%%%%%%%%%%%%%%%%%%%%%%%%%%%%%%%%%%%

A standard convolution of the cross section (\ref{result}) with the gluon distribution in the nucleon gives the cross section of inclusive vector leptoquark production measurable in electron--proton collisions:  

\begin{equation}
\sigma_{e_Lp\rightarrow VX}(s,M_V,m_q)=\int_{x_0}^1dx\,\text{g}(x,s)\sigma_2(xs,M_V,m_q),\label{convolution}
\end{equation}

where $\text{g}(x,s)$ is the gluon distribution function,
$x_0=(M_V+m_q)^2/s$. 

\begin{figure}
\centering
\resizebox{0.9\textwidth}{!}{%
\includegraphics{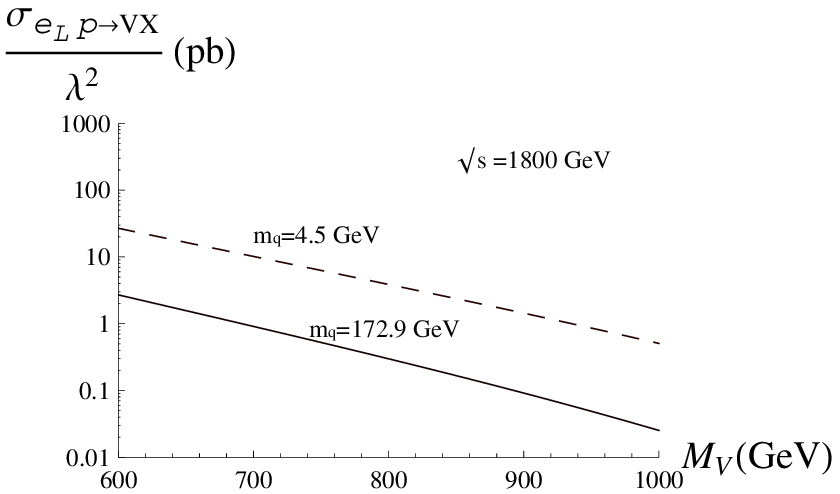}
}\caption{}
\label{fig7}
\end{figure}

Figure~\ref{fig7} shows the cross sections for the production of the ($eb$) and ($et$) type vector leptoquarks evaluated using (\ref{convolution}) with the gluon distribution function adopted from CTEQ5~\cite{gluon_distr}.

%%%%%%%%%%%%%%%%%%%%%%%%%%%%%%%%%%%%
\section{Conclusions}
%%%%%%%%%%%%%%%%%%%%%%%%%%%%%%%%%%%%%
The cross section for vector leptoquark production in electron--gluon collisions is calculated analytically using the $SU(3)_C\times SU(2)_L\times U(1)_Y$-symmetric Lagrangian with the minimal couplings between the leptoquarks and the gauge fields of the standard model. It is shown that from a formal point of view this process is similar to  exclusive production of $W$ bosons in neutrino--photon scattering allowed by the standard model whose cross section is also found.  The cross section of the leptoquark production obtained in this letter significantly exceeds the corresponding result of previous calculations presented in the literature. The cross sections of inclusive production of the ($eb$) and ($et$) vector leptoquarks in electron--nucleon interactions observable at $ep$ colliders are evaluated. The presented analysis is  applicable to studies of the production of vector leptoquarks of the other types as well.

%%%%%%%%%%%%%%%%%%%%  
{\bf Acknowledgements}
%%%%%%%%%%%%%%%%%%%
\vskip 0.5cm

I thank P.M. Zerwas for useful comments on production of leptoquarks in several processes.
I am also thankful to M. Spira for drawing my attention to their results on production of leptoquarks at $ep$ colliders as well as for providing useful information. This work was supported in part by the Russian Foundation for Basic Research (grant 11-02-12043), by the Program for Basic Research of the Presidium of the Russian Academy of Sciences "Neutrino Physics and Neutrino Astrophysics" and by the Federal Target Program  of the Ministry of Education and Science of Russian Federation "Research and Development in Top Priority Spheres of Russian Scientific and Technological Complex for 2007-2013" (contract No. 16.518.11.7072).

%%%%%%%%%%%%%%%%%%%%%%%%%

\newpage

%%%%%%%%%%%%%%%%%%%%%
{\bf Figure Captions}
%%%%%%%%%%%%%%%%%%%%%
\vskip 0.5 cm 

{\bf Fig. 1:} The Feynman rules: (a) for the $\gamma WW$ vertex; (b) for the $\text{g}VV$ vertex.

{\bf Fig. 2:} Tree level Feynman diagrams describing the process $\nu_{l}+\gamma\rightarrow l+W$.

\vskip 0.5 cm

{\bf Fig. 3:} Tree level Feynman diagrams describing the process $e+\text{g}\rightarrow q+V$. 

\vskip 0.5 cm

{\bf Fig. 4:} The Feynman rules: (a) for the $W\nu l$ vertex; (b) for the $Veq$ vertex.

\vskip 0.5 cm

{\bf Fig. 5:} Dependence of the ratio of the cross section $\sigma_1^c(s,M_W,m_l)$ taken from~\cite{seckel} to that calculated in the present letter on the cm energy in the resonance region. 

\vskip 0.5 cm

{\bf Fig. 6:} Dependence of the cross sections of production of vector leptoquarks of mass 1000 GeV on the cm energy. Right panel: the $(eb)$ type leptoquark. Left panel: the $(et)$ type leptoquark. The solid curves represent the results of this letter, the dashed curves are the calculations of \cite{spira}. The couplings $\lambda=1$ and $\alpha_s=0.118$.

\vskip 0.5 cm

{\bf Fig. 7:} Dependence of the ratio of the cross section $\sigma_2^c(s,M_V,m_q)$ taken from~\cite{spira} to that calculated in the present letter on the cm energy for production of the $(et)$ type vector leptoquark of mass 500 GeV (dash-dotted), 700 GeV (dashed) and 1000 GeV (solid). The dotted line corresponds to $\sigma^c_2/\sigma_2=1$. 

\vskip 0.5 cm

{\bf Fig. 8:} The cross section of the reaction $e_Lp\rightarrow VX$ as a function of the leptoquark mass at $\sqrt{s}=1800$ GeV. Dashed curve: the $(eb)$ type leptoquark. Solid curve: the $(et)$ type leptoquark. The coupling $\lambda$ is divided out, $\alpha_s=0.118$. 

%%%%%%%%%%%%%%%%%%%%%

\newpage
%%%%%%%%%%
%%%Figures
%%%%%%%%%%

%%%%%%%%%%%%%%
% FIGURE 1
%%%%%%%%%%%%%%
%%%%%%%%%%%%%%%%%%%%

%%%%%%%%%%%%%%
% FIGURE 2
%%%%%%%%%%%%%%
%%%%%%%%%%%%%%%%%%%%

%%%%%%%%%%%%%%
% FIGURE 3
%%%%%%%%%%%%%%
%%%%%%%%%%%%%%%%%%%%

%%%%%%%%%%%%%%
% FIGURE 4
%%%%%%%%%%%%%%
%%%%%%%%%%%%%%%%%%%%

%%%%%%%%%%%%%%
% FIGURE 5
%%%%%%%%%%%%%%

%%%%%%%%%%%%%%
% FIGURE 6
%%%%%%%%%%%%%%

%%%%%%%%%%%%%%
% FIGURE 7
%%%%%%%%%%%%%%

%%%%%%%%%%%%%%
% FIGURE 8
%%%%%%%%%%%%%%

\end{document}